\documentclass[11pt]{article}
\def\s{\sigma}

\begin{document}
\title{{\bf{\Large Strings in pp-wave background and background
B-field from membrane and its symplectic quantization}}}
\author{
 {\bf {\normalsize Sunandan Gan{g}opadhyay}\thanks{sunandan@bose.res.in}} \\
  {\normalsize S.~N.~Bose National Centre for Basic Sciences,}\\{\normalsize JD Block, Sector III, Salt Lake, Kolkata-700098, India}\\[0.3cm]
}

\maketitle

\begin{abstract}
\noindent The symplectic quantization technique is applied to
open free membrane and strings in pp-wave background and
background gauge field obtained by compactifying
the open membrane in the presence of a background
anti-symmetric 3--form field. In both cases, 
first the Poisson brackets among the Fourier modes 
are obtained and then the Poisson brackets among
the membrane(string) coordinates are computed.
The full noncommutative phase-space structure is reproduced in case
of strings in pp-wave background and background
gauge field. We feel that this method of obtaining the
Poisson algebra is more elegant than previous 
approaches discussed in the literature. 

\\[0.3cm]

\end{abstract}

\section{Introduction}
It has been discovered recently that there exists a new maximal
supersymmetric IIB supergravity background, namely pp-wave
Ramond-Ramond (RR) background \cite{blau}. It consists of a plane-wave
metric supported by a homogeneous RR 5--form flux
\begin{eqnarray}
\label{intro}
ds^{2}&=&2dX^{+}dX^{-}-\mu^{2}X^{I}X^{I}(dX^{+})^{2}+dX^{I}dX^{I}
\quad,\quad I=1,...,8\nonumber\\
&&F_{+1234}=F_{+5678}=2\mu~.
\end{eqnarray}
The background has 32 symmetries and is related (by a special limit
\cite{pen}) to the $AdS_{5}\times S^{5}$ background 
\cite{blau, hull, papa}. Remarkably,
string theory in this background is exactly solvable \cite{met}. Solvability
in this context means that it is possible to find exact solutions of the
classical string equations of motion, perform a canonical quantization
and then determine the Hamiltonian operator.

\noindent On the other hand, the theory of membranes has also been
studied extensively over the last decade \cite{wt}. In a recent paper
\cite{kamani}, it has been shown that one can obtain the action of an infinite
number of massive strings in the pp-wave background by compactifying
the bosonic membrane action. Some properties of closed and open strings
in this background has also been investigated in this paper.

\noindent Noncommutativity is another area which has attracted a lot
of attention in the past few years \cite{sw} owing to the inspiration
of superstring theories. It is a well known result now that open strings
attached to D--branes in the presence of a background 
anti-symmetric B--field induces 
noncommutativity at its end points, 
i.e. along the world--volume of 
the D--brane \cite{ho, ho1, ard}. This fact further
reveals the nontrivial role of boundary conditions (BC) in string theory
and the need of taking them into account when considering
the quantization of open strings. The most conventional way in which
this result has been derived is by employing the 
Dirac approach \cite{dir} with the string BC(s) imposed as  
second class constraints in \cite{ar, ho1, ar11, br1}. 
A more elegant approach
of obtaining this noncommutativity, 
done in spirit to the treatment of Hanson et.al 
\cite{hrt} (where modified Poisson brackets 
(PB) were obtained for the free Nambu-Goto string), 
is by modifying the canonical bracket structure, 
so that it is compatible with the BC(s) 
\cite{rb, rbbckk, bcsgaghs, sgaghas}. A conformal field
theoretic approach to this problem
has been studied in \cite{bcsgag, sgag}. In \cite{jian}, the
Faddeev-Jackiw (FJ) symplectic formalism \cite{faddeev}  
has been applied to obtain the PB(s) among the Fourier modes
which appear in the solutions of the 
classical string equations of motion.  
Using this they obtained the 
PB(s) among the open string coordinates revealing 
the noncommutative structure at the string end points.

\noindent In this paper, we first show that strings in pp-wave background
and background gauge fields can be obtained by compactifying the open
membrane action in the presence of a background 3--form anti-symmetric
field. We then employ the FJ technique (as done before in \cite{jian}) 
to obtain the PB(s) among the Fourier modes and then 
between the string coordinates. The advantage of this method is that
one need not know all the constraint chains by the consistency
requirements that arise in the Dirac approach.

\noindent The organization of this paper is as follows. To illustrate
our method, we begin by discussing the free open membrane in section 2.
In section 3, we discuss the interacting membrane and the
symplectic structure of strings in pp-wave
background and background gauge field obtained by compactifying the
interacting membrane. Finally, we conclude in section 4.

\section{Free open membrane}
An open membrane is a two dimensional object which
sweeps out a three dimensional world-volume parametrized
by $\tau$, $\sigma^{1}$ and $\sigma^{2}$. These parameters
can be collectively referred to as $\xi_{i}$, $(i=0, 1, 2)$.
The Polyakov action for the bosonic membrane is then given by \cite{wt}
\begin{eqnarray}
S&=&\int d\tau~L =-\frac{1}{4\pi\alpha^{\prime}}\int 
d^{3}\xi~\left(g^{ab}\eta_{\mu\nu}\partial_{a}X^{\mu}
\partial_{b}X_{\mu} - 1\right)\quad;\quad (a, b= 0, 1, 2)
\label{1}
\end{eqnarray}
where, 
\begin{eqnarray}
L&=&-\frac{1}{4\pi\alpha^{\prime}}~\int_{0}^{\pi}\int_{0}^{\pi}~
d\sigma^{1}d\sigma^{2}\left(g^{ab}
\eta_{\mu\nu}\partial_{a}X^{\mu}
\partial_{b}X_{\mu} - 1\right)
\label{1a}
\end{eqnarray}
is the Lagrangian and $g_{ab}=diag(-, +, +)$,
$\eta_{\mu\nu}=diag(-, +,..., +)$. The final term $(-1)$
stands for the cosmological constant and does not appear
in the string theory action.

\noindent The variation of (\ref{1}) gives the equation of motion
\begin{equation}
\label{2}
(\partial^{2}_{0}-\partial^{2}_{1}
-\partial^{2}_{2})X^{\mu}(\tau, \sigma^{1}, \sigma^{2})=0
\end{equation}
and two types of BC(s). They are the Dirichlet BC(s)
\begin{eqnarray}
\label{3}
\delta X^{\mu}(\tau, \sigma^{1}, \sigma^{2})
|_{\sigma^{1}=0, \pi}&=&0 \nonumber\\
\delta X^{\mu}(\tau, \sigma^{1}, \sigma^{2})
|_{\sigma^{2}=0, \pi}&=&0
\end{eqnarray}
and the Neumann BC(s)
\begin{eqnarray}
\label{4}
\partial_{1}X^{\mu}(\tau, \sigma^{1}, \sigma^{2})
|_{\sigma^{1}=0, \pi}&=&0\nonumber\\
\partial_{2}X^{\mu}(\tau, \sigma^{1}, \sigma^{2})
|_{\sigma^{2}=0, \pi}&=&0~.
\end{eqnarray}
The canonically conjugate momenta $\Pi_{\mu}$ to $X^{\mu}$ reads
(we set $2\pi \alpha^\prime=1$ for convenience and we shall recover
it whenever neccessary):
\begin{eqnarray}
\label{momenta}
\Pi_{\mu}(\tau, \sigma^{1}, \sigma^{2})&=&
\frac{\delta S_{P}}{\delta(\partial_{0}
X^{\mu}(\tau, \sigma^{1}, \sigma^{2}))}=\eta_{\mu\nu}\partial_{0}
X^{\nu}.
\end{eqnarray}
In order to quantize consistently, we need well-defined PB(s)
among the canonical variables 
$X^{\mu}(\tau, \sigma^{1}, \sigma^{2})$ 
and $\Pi_{\mu}(\tau, \sigma^{1}, \sigma^{2})$ 
which read:
\begin{eqnarray}
\label{Poisson}
\{X^{\mu}(\tau, \sigma^{1}, \sigma^{2}),~\Pi_{\nu}(\tau, \sigma^{1'}, 
\sigma^{2'})\}&=&\delta^{\mu}_{\nu}~\delta(\sigma^{1}-\sigma^{1'})
~\delta(\sigma^{2}-\sigma^{2'})
\end{eqnarray}
\begin{eqnarray}
\label{Poisson1}
\{X^{\mu}(\tau, \sigma^{1}, \sigma^{2}),~X^{\nu}(\tau, \sigma^{1'}, 
\sigma^{2'})\}&=&
\{\Pi_{\mu}(\tau, \sigma^{1}, \sigma^{2}),~ \Pi_{\nu}(\tau, \sigma^{1'}, 
\sigma^{2'})\}=0~.
\end{eqnarray}
The Hamiltonian is obtained by means
of Legendre transformation
\begin{eqnarray}
H&=&\int_{0}^{\pi}\int_{0}^{\pi}d\sigma^{1}d\sigma^{2}~\Pi_{\mu}
\partial_{0}X^{\mu}-L\nonumber\\
&=&\frac{1}{2}\int_{0}^{\pi}\int_{0}^{\pi}d\sigma^{1}d\sigma^{2}
\left[\eta_{\mu\nu}\left(\Pi^{\mu}\Pi^{\nu}+\partial_{1}X^{\mu}
\partial_{1}X^{\nu}+\partial_{2}X^{\mu}
\partial_{2}X^{\nu}\right)-1\right]
\label{hamil}
\end{eqnarray}
and the time-evolution of $X^{\mu}(\tau, \sigma^{1}, \sigma^{2})$,
$\Pi_{\mu}(\tau, \sigma^{1}, \sigma^{2})$ is governed by
\begin{eqnarray}
\partial_{0}X^{\mu}(\tau, \sigma^{1}, \sigma^{2})
&=&\{X^{\mu}(\tau, \sigma^{1}, \sigma^{2}),~H\}
\label{evo}
\end{eqnarray}
\begin{eqnarray}
\partial_{0}\Pi_{\mu}(\tau, \sigma^{1}, \sigma^{2})
&=&\{\Pi_{\mu}(\tau, \sigma^{1}, \sigma^{2}),~H\}.
\label{evo1}
\end{eqnarray}
However, at the open membrane end points, the
PB(s) (\ref{Poisson}) are not compatible with
the BC(s) (\ref{3}, \ref{4}). This implies that
the basic PB(s) must be modified in order to make 
them consistent with the BC(s). In the rest of the paper, we 
shall work with the Neumann BC(s) (\ref{4}).

\noindent The solution to the equations of motion (\ref{2})
compatible with the BC(s) (\ref{4}) read
\begin{eqnarray}
\label{sol}
X^{\mu}(\tau, \sigma^{1}, \sigma^{2})
&=& x^{\mu}_{0}+p^{\mu}\tau+i\sum_{n=1}^{\infty}\frac{1}{\sqrt{n}}
\left(\alpha^{\mu}_{n0}~e^{in\tau}-\alpha^{\mu~\dagger}_{n0}
~e^{-in\tau}\right)
\cos(n\sigma^{1})\nonumber\\
&& + i\sum_{m=1}^{\infty}\frac{1}{\sqrt{m}}
\left(\alpha^{\mu}_{0m}~e^{im\tau}-\alpha^{\mu~\dagger}_{0m}~
e^{-im\tau}\right)\cos(m\sigma^{2})\nonumber\\
&& +i\sum_{n, m=1}^{\infty}(n^2+m^2)^{-1/4}
\left(\alpha^{\mu}_{nm}~e^{i\sqrt{n^2+m^2}\tau}-
\alpha^{\mu~\dagger}_{nm}~e^{-i\sqrt{n^2+m^2}\tau}\right)
\cos(n\sigma^{1})\cos(m\sigma^{2})\nonumber\\
&=& x^{\mu}_{0}+p^{\mu}\tau+i\sum_{n=1}^{\infty}\frac{1}{\sqrt{n}}
\left(\alpha^{\mu}_{n0}(\tau)-\alpha^{\mu~\dagger}_{n0}
(\tau)\right)\cos(n\sigma^{1})\nonumber\\
&& + i\sum_{m=1}^{\infty}\frac{1}{\sqrt{m}}
\left(\alpha^{\mu}_{0m}(\tau)-\alpha^{\mu~\dagger}_{0m}(\tau)
\right)\cos(m\sigma^{2})\nonumber\\
&& +i\sum_{n, m=1}^{\infty}(n^2+m^2)^{-1/4}
\left(\alpha^{\mu}_{nm}(\tau)-
\alpha^{\mu~\dagger}_{nm}(\tau)\right)\cos(n\sigma^{1})\cos(m\sigma^{2})
\end{eqnarray}
where we have defined 
$\alpha^{\mu}_{n0}(\tau)=\alpha^{\mu}_{n0}~e^{in\tau}$,
$\alpha^{\mu~\dagger}_{n0}(\tau)=\alpha^{\mu~\dagger}_{n0}~e^{-in\tau}$, 
and so on.\\

\noindent The canonically conjugate momenta 
(\ref{momenta}) expressed in terms of the Fourier components read

\begin{eqnarray}
\label{mom}
\Pi_{\mu}(\tau, \sigma^{1}, \sigma^{2})
=&&\eta_{\mu\nu}\left[p^{\nu}-\sum_{n=1}^{\infty}\sqrt{n}
\left(\alpha^{\nu}_{n0}(\tau)+\alpha^{\nu~\dagger}_{n0}(\tau)\right)
\cos(n\sigma^{1})\right.\nonumber\\
&&\left. -\sum_{m=1}^{\infty}\sqrt{m}
\left(\alpha^{\nu}_{0m}(\tau)+\alpha^{\nu~\dagger}_{0m}(\tau)\right)
\cos(m\sigma^{2})\right.\nonumber\\
&&\left. -\sum_{n, m=1}^{\infty}(n^2+m^2)^{1/4}
\left(\alpha^{\nu}_{nm}(\tau)+
\alpha^{\nu~\dagger}_{nm}(\tau)\right)
\cos(n\sigma^{1})\cos(m\sigma^{2})\right].\nonumber\\
\end{eqnarray}
We shall now use the FJ method \cite{faddeev}
to obtain the PB(s) between the Fourier components.
The idea is to write a Lagrangian in the first-order form
\begin{eqnarray}
\label{first}
L&=& a_{n}(\xi)\partial_{0}\xi^{n}-H
\end{eqnarray}
where $\xi^{n}$ stand for all the canonical variables
and $a_{n}(\xi)$ can be read directly from the inverse
of the matrix
\begin{eqnarray}
f_{mn}&=&\frac{\partial a_{n}(\xi)}{\partial\xi^{m}}
-\frac{\partial a_{m}(\xi)}{\partial\xi^{n}}
\label{mat1}
\end{eqnarray}
provided the inverse of $f_{mn}$ exists.
The first order form of the Lagrangian (\ref{1a}) reads:
\begin{eqnarray}
\label{first1}
L&=& \int_{0}^{\pi}\int_{0}^{\pi}
d\sigma^{1}d\sigma^{2}~\Pi_{\mu}\partial_{0}X^{\mu}-H~.
\end{eqnarray}
Substituting (\ref{sol}) and (\ref{mom}) in the above equation
yields
\begin{eqnarray}
\label{lagrangian}
L=&&\pi^{2}\eta_{\mu\nu}\left[p^{\mu}p^{\nu}-\frac{i}{2}\sum_{n=1}^{\infty}
\left(\alpha^{\nu}_{n0}(\tau)+\alpha^{\nu~\dagger}_{n0}(\tau)\right)
\left(\dot{\alpha}^{\mu}_{n0}(\tau)-\dot{\alpha}^{\mu~\dagger}_{n0}
(\tau)\right)
\right.\nonumber\\
&&\left. -\frac{i}{2}\sum_{m=1}^{\infty}
\left(\alpha^{\nu}_{0m}(\tau)+\alpha^{\nu~\dagger}_{0m}(\tau)\right)
\left(\dot{\alpha}^{\mu}_{0m}(\tau)-\dot{\alpha}^{\mu~\dagger}_{0m}
(\tau)\right)
\right.\nonumber\\
&&\left. -\frac{i}{4}\sum_{n, m=1}^{\infty}
\left(\alpha^{\nu}_{nm}(\tau)+
\alpha^{\nu~\dagger}_{nm}(\tau)\right)
\left(\dot{\alpha}^{\mu}_{nm}(\tau)-\dot{\alpha}^{\mu~\dagger}_{nm}
(\tau)\right)
\right]-H\nonumber\\
\end{eqnarray}
where, 
\begin{eqnarray}
\label{ham}
H=&&\frac{\pi^{2}}{2}\eta_{\mu\nu}\left[p^{\mu}p^{\nu}+\sum_{n=1}^{\infty}
n\left(\alpha^{\mu~\dagger}_{n0}~\alpha^{\nu}_{n0}+
\alpha^{\nu~\dagger}_{n0}~\alpha^{\mu}_{n0}\right)
\right.\nonumber\\
&&\left. +\sum_{m=1}^{\infty}
m\left(\alpha^{\mu~\dagger}_{0m}~\alpha^{\nu}_{0m}+
\alpha^{\nu~\dagger}_{0m}~\alpha^{\mu}_{0m}\right)
\right.\nonumber\\
&&\left. +\frac{1}{2}\sum_{n, m=1}^{\infty}
(n^2+m^2)^{1/2}\left(\alpha^{\mu~\dagger}_{nm}~\alpha^{\nu}_{nm}+
\alpha^{\nu~\dagger}_{nm}~\alpha^{\mu}_{nm}\right)\right]-\frac{\pi^2}{2}
\nonumber\\
\end{eqnarray}
and the dots denote differentiation w.r.t. $\tau$.

\noindent It is now easy to read from the above first order form of the
Lagrangian (\ref{lagrangian}) three sets of variables
$\xi^{n}=(\alpha^{\mu}_{n0}(\tau), \alpha^{\mu~\dagger}_{n0}(\tau))$,
$\xi^{m}=(\alpha^{\mu}_{0m}(\tau), \alpha^{\mu~\dagger}_{0m}(\tau))$,
$\xi^{nm}=(\alpha^{\mu}_{nm}(\tau), \alpha^{\mu~\dagger}_{nm}(\tau))$,
and their corresponding one forms 
$a_{n}(\xi)=-\frac{i\pi^2}{2}\eta_{\mu\nu}[(\alpha^{\nu}_{n0}(\tau)+
\alpha^{\nu~\dagger}_{n0}(\tau)), -(\alpha^{\nu}_{n0}(\tau)+
\alpha^{\nu~\dagger}_{n0}(\tau))]$,
$a_{m}(\xi)=-\frac{i\pi^2}{2}\eta_{\mu\nu}[(\alpha^{\nu}_{0m}(\tau)+
\alpha^{\nu~\dagger}_{0m}(\tau)), -(\alpha^{\nu}_{0m}(\tau)+
\alpha^{\nu~\dagger}_{0m}(\tau))]$, and
$a_{nm}(\xi)=-\frac{i\pi^2}{4}\eta_{\mu\nu}[(\alpha^{\nu}_{nm}(\tau)+
\alpha^{\nu~\dagger}_{nm}(\tau)), -(\alpha^{\nu}_{nm}(\tau)+
\alpha^{\nu~\dagger}_{nm}(\tau))]$. The matrix $f$ for these
three sets of variables can now be computed using (\ref{mat1})
and the result is
\begin{eqnarray}
f&=& \pmatrix{0&B\cr
-B&0\cr}
\label{mat2}
\end{eqnarray}
in which $0$ is a null matrix and B is a diagonal matrix
\begin{eqnarray}
B^{\mu\nu}_{nn^{'}}&=&i\pi^{2}\eta^{\mu\nu}~\delta_{nn^{'}}\quad;
\quad (n, n^{'}=1,2,...)\quad for\quad \alpha^{\mu}_{n0}~~ modes
\label{mat3}
\end{eqnarray}
\begin{eqnarray}
B^{\mu\nu}_{mm^{'}}&=&i\pi^{2}\eta^{\mu\nu}~\delta_{mm^{'}}\quad;
\quad (m, m^{'}=1,2,...)\quad for\quad \alpha^{\mu}_{0m}~~ modes
\label{mat4}
\end{eqnarray}
\begin{eqnarray}
B^{\mu\nu}_{nn^{'}mm^{'}}&=&\frac{i\pi^{2}}{2}\eta^{\mu\nu}~\delta_{nn^{'}}~
\delta_{mm^{'}}\quad;\quad(n, n^{'}, m, m^{'}=1,2,...)
\quad for\quad \alpha^{\mu}_{nm}~~ modes.
\label{mat5}
\end{eqnarray}
The inverse of the matrix $f$ can be easily obtained and reads
\begin{eqnarray}
f^{-1}&=& \pmatrix{0&-\frac{1}{B}\cr
\frac{1}{B}&0\cr}.
\label{mat6}
\end{eqnarray}
Hence, according to FJ method, the non-trivial PB(s) are given by
\begin{eqnarray}
\{\alpha^{\mu}_{n0}(\tau),~\alpha^{\nu~\dagger}_{n^{'}0}(\tau)\}
&=&\frac{i}{\pi^2}\eta^{\mu\nu}~\delta_{nn^{'}}\nonumber\\
\{\alpha^{\mu}_{0m}(\tau),~\alpha^{\nu~\dagger}_{0m^{'}}(\tau)\}
&=&\frac{i}{\pi^2}\eta^{\mu\nu}~\delta_{mm^{'}}\nonumber\\
\{\alpha^{\mu}_{nm}(\tau),~\alpha^{\nu~\dagger}_{n^{'}m^{'}}(\tau)\}
&=&\frac{2i}{\pi^2}\eta^{\mu\nu}~\delta_{nn^{'}}~\delta_{mm^{'}}
\label{pb1}
\end{eqnarray}
which further reduces to
\begin{eqnarray}
\{\alpha^{\mu}_{n0},~\alpha^{\nu~\dagger}_{n^{'}0}\}
&=&\frac{i}{\pi^2}\eta^{\mu\nu}~\delta_{nn^{'}}\nonumber\\
\{\alpha^{\mu}_{0m},~\alpha^{\nu~\dagger}_{0m^{'}}\}
&=&\frac{i}{\pi^2}\eta^{\mu\nu}~\delta_{mm^{'}}\nonumber\\
\{\alpha^{\mu}_{nm},~\alpha^{\nu~\dagger}_{n^{'}m^{'}}\}
&=&\frac{2i}{\pi^2}\eta^{\mu\nu}~\delta_{nn^{'}}~\delta_{mm^{'}}.
\label{pb2}
\end{eqnarray}
Now substituting (\ref{sol}) and (\ref{ham}) 
in (\ref{evo}) leads to the PB(s) among the zero modes:
\begin{eqnarray}
\{x^{\mu}_{0},~p^{\nu}\}&=&\frac{1}{\pi^2}\eta^{\mu\nu}.
\label{pb3}
\end{eqnarray}
With the above results in hand, the PB(s) among the canonical variables
$X^{\mu}(\tau, \sigma^{1}, \sigma^{2})$ and 
$\Pi_{\mu}(\tau, \sigma^{1}, \sigma^{2})$ read:
\begin{eqnarray}
\{X^{\mu}(\tau, \sigma^{1}, \sigma^{2}),~
X^{\nu}(\tau, \sigma^{1'}, \sigma^{2'})\}&=&
\{\Pi_{\mu}(\tau, \sigma^{1}, \sigma^{2}),~
\Pi_{\nu}(\tau, \sigma^{1'}, \sigma^{2'})\}=0
\label{pb5}
\end{eqnarray}
\begin{eqnarray}
\{X^{\mu}(\tau, \sigma^{1}, \sigma^{2}),~
\Pi_{\nu}(\tau, \sigma^{1'}, \sigma^{2'})\}
&=&\delta^{\mu}_{\nu}~\Delta_{+}(\sigma^{1},~\sigma^{1'})
~\Delta_{+}(\sigma^{2},~\sigma^{2'})
\label{pb4}
\end{eqnarray}
where,
\begin{eqnarray}
\label{delta}
\Delta_{+}(\s ,\s' )={1\over \pi }\left(1 +
\sum_{n \neq 0}\cos(n\sigma)\cos(n\sigma^{\prime})\right)
\end{eqnarray}
satisfies the usual properties of the delta
function in the interval $[0, \pi]$ \cite{hrt}.

\noindent Note that the above symplectic structure is consistent
with the Neumann BC(s) (\ref{4}). 

\section{Open membrane in the constant three-form field background}
The Polyakov action of a membrane in the presence of a 
background anti-symmetric three-form field $A_{\mu\nu\rho}$ reads:
\begin{eqnarray}
S&=&-\frac{1}{4\pi\alpha^{\prime}}\int 
d^{3}\xi~\left[\left(g^{ab}\eta_{\mu\nu}\partial_{a}X^{\mu}
\partial_{b}X_{\mu} - 1\right)+\frac{1}{3}\epsilon^{abc}
A_{\mu\nu\rho}\partial_{a}X^{\mu}\partial_{b}X^{\nu}
\partial_{c}X^{\rho}\right].
\label{int1}
\end{eqnarray}
where $\epsilon^{abc}$ is anti-symmetric in all the indices.

\noindent The variation of the above action leads to the equations
of motion (\ref{2}) and the BC(s)
\begin{eqnarray}
\left(\partial_{1}X^{\mu}-{A^{\mu}}_{\nu\rho}\partial_{0}X^{\nu}
\partial_{2}X^{\rho}\right)|_{\sigma^{1}=0, \pi}&=&0
\label{int2}
\end{eqnarray}
\begin{eqnarray}
\left(\partial_{2}X^{\mu}+{A^{\mu}}_{\nu\rho}\partial_{0}X^{\nu}
\partial_{1}X^{\rho}\right)|_{\sigma^{2}=0, \pi}&=&0~.
\label{int3}
\end{eqnarray}
The canonically conjugate momenta $\Pi_{\mu}$ to $X^{\mu}$ is given by:
\begin{eqnarray}
\Pi_{\mu}(\tau, \sigma^{1}, \sigma^{2})
&=&\eta_{\mu\nu}\partial_{0}X^{\nu}-A_{\mu\nu\rho}
\partial_{1}X^{\nu}\partial_{2}X^{\rho}~.
\label{int4}
\end{eqnarray}
Using $\Pi_{\mu}$, the BC(s) (\ref{int2}), (\ref{int3}) can be expressed
in terms of the phase-space variables as:
\begin{eqnarray}
\left[\left({\eta^{\mu}}_{\kappa}-{A^{\mu}}_{\nu\rho}
{A^{\nu}}_{\kappa\beta}\partial_{2}X^{\beta}\partial_{2}X^{\rho}\right)
\partial_{1}X^{\kappa}
-{A^{\mu}}_{\nu\rho}\Pi^{\nu}\partial_{2}X^{\rho}\right]
|_{\sigma^{1}=0, \pi}&=&0
\label{int5}
\end{eqnarray}
\begin{eqnarray}
\left[\left({\eta^{\mu}}_{\beta}+{A^{\mu}}_{\nu\rho}
{A^{\nu}}_{\kappa\beta}\partial_{1}X^{\kappa}\partial_{1}X^{\rho}\right)
\partial_{2}X^{\beta}
+{A^{\mu}}_{\nu\rho}\Pi^{\nu}\partial_{1}X^{\rho}\right]
|_{\sigma^{2}=0, \pi}&=&0~.
\label{int6}
\end{eqnarray}
The above form of the BC(s) indicates that it is problematic to
find exact solutions to the equations of motion (\ref{2}).
So we study the low energy limit where the membrane goes
to string theory in the limit of small radius for
the cylindrical membrane.

\noindent To do this, we take the $\sigma^{2}$-- direction of the membrane
to be wrapped around a circle with radius R.
\noindent We choose further the 
gauge fixing condition \cite{kamani, rbbckk}\footnote{Note that
if the $\sigma^{2}$-- direction of the membrane
is wrapped around a circle of radius R, then the $X^{2}$-direction
is also compact on the same circle.}
\begin{equation}
\quad X^{2} = \sigma^{2}\quad;\quad (0\leq\sigma^{2}\leq 2\pi R).
\label{int7}
\end{equation}
Now substituting the Fourier expansion of the world-volume fields
$X^{\mu}(\tau, \sigma^{1}, \sigma^{2})$ ($\mu\neq2$):
\begin{eqnarray}
X^{\mu}(\tau, \sigma^{1}, \sigma^{2})
&=&\sum_{n=-\infty}^{+\infty}X^{\mu}_{n}(\tau, \sigma^{1})
~e^{in\sigma^{2}/R}\quad;\quad 
X^{\mu}_{-n}(\tau, \sigma^{1})=X^{\mu~\dagger}_{n}(\tau, \sigma^{1})
\label{int8}
\end{eqnarray} 
in the action (\ref{int1}) and using (\ref{int7}), we obtain 
(recovering the $2\pi\alpha^{\prime}$ factor):
\begin{eqnarray}
S=&&\frac{1}{2\tilde\alpha^{\prime}}
\sum_{n=-\infty}^{+\infty}\int d\tau d\sigma^{1}
\left(\partial_{0}X^{\mu}_{n}(\tau, \sigma^{1})
\partial_{0}X_{-n\mu}(\tau, \sigma^{1})
-\partial_{1}X^{\mu}_{n}(\tau, \sigma^{1})
\partial_{1}X_{-n\mu}(\tau, \sigma^{1})
\right.\nonumber\\
&&~~~~~~~~~~~~~~~\left.-m_{n}^{2}X^{\mu}_{n}(\tau, \sigma^{1})
X_{-n\mu}(\tau, \sigma^{1})-A_{\mu\nu2}
~\partial_{0}X^{\mu}_{n}(\tau, \sigma^{1})
\partial_{1}X^{\nu}_{-n}(\tau, \sigma^{1})\right)\nonumber\\
&&+\frac{1}{2\tilde\alpha^{\prime}}
\sum_{n=m\neq0, n\neq-m}\frac{i(n+m)}{R}A_{\mu\nu\rho\neq2}
~\partial_{0}X^{\mu}_{n}(\tau, \sigma^{1})
\partial_{1}X^{\nu}_{m}(\tau, \sigma^{1})X^{\rho\neq2}_{-(n+m)}
(\tau, \sigma^{1});~~ 
m_{n}=\frac{|n|}{R}~,~\tilde\alpha^{\prime}=\alpha^{\prime}/R\nonumber\\
=&&\frac{1}{2\tilde\alpha^{\prime}}
\left[S_{0}+\sum_{n\neq0}S_{n}+\sum_{n=m\neq0, n\neq-m}
\frac{i(n+m)}{R}\int d\tau d\sigma^{1}A_{\mu\nu\rho\neq2}
~\partial_{0}X^{\mu}_{n}(\tau, \sigma^{1})
\partial_{1}X^{\nu}_{m}(\tau, \sigma^{1})
X^{\rho\neq2}_{-(n+m)}
(\tau, \sigma^{1})\right]\nonumber\\
\label{int9}
\end{eqnarray}
where,
\begin{eqnarray}
S_{0}=&&\frac{1}{2\tilde\alpha^{\prime}}
\int d\tau d\sigma^{1}
\left[\dot{X}^{\mu}_{0}(\tau, \sigma^{1})
\dot{X}_{0\mu}(\tau, \sigma^{1})
-\partial_{1}X^{\mu}_{0}(\tau, \sigma^{1})
\partial_{1}X_{0\mu}(\tau, \sigma^{1})\right.\nonumber\\
&&\left.~~~~~~~~~~~~~~~~~~~~~~~~~~~~~~~~~-B_{\mu\nu}
~\dot{X}^{\mu}_{0}(\tau, \sigma^{1})
\partial_{1}X^{\nu}_{0}(\tau, \sigma^{1})\right]
\label{int10}
\end{eqnarray}
is the usual low energy string theory action in the presence
of background gauge field $A_{\mu\nu2}=B_{\mu\nu}$ and
\begin{eqnarray}
S_{n}=&&\frac{1}{2\tilde\alpha^{\prime}}
\int d\tau d\sigma^{1}
\left[\partial_{0}X^{\mu}_{n}(\tau, \sigma^{1})
\partial_{0}X_{-n\mu}(\tau, \sigma^{1})
-\partial_{1}X^{\mu}_{n}(\tau, \sigma^{1})
\partial_{1}X_{-n\mu}(\tau, \sigma^{1})\right.\nonumber\\
&&\left.-m_{n}^{2}X^{\mu}_{n}(\tau, \sigma^{1})
X_{-n\mu}(\tau, \sigma^{1})-B_{\mu\nu}
~\partial_{0}X^{\mu}_{n}(\tau, \sigma^{1})
\partial_{1}X^{\nu}_{-n}(\tau, \sigma^{1})\right]
\label{int11}
\end{eqnarray}
is the action of massive strings in pp-wave background
and background gauge field $B_{\mu\nu}$ \cite{chu}.

\noindent Clearly, from (\ref{int9}), (\ref{int10}) and (\ref{int11}) 
we observe that the last term contains 
modes which are higher in energy
than the first two terms. Hence in the low energy limit, we only
consider the first two terms in the action (\ref{int9}).
The symplectic quantization of the usual string theory
action (\ref{int10}) leads to the well-known noncommutativity
at the end points of the string \cite{jian}. 
In this paper, we shall carry
out the symplectic quantization of 
massive strings in pp-wave background
and background gauge field $B_{\mu\nu}$.

\noindent The variation of $S_{n}$ gives the equation
of motion
\begin{eqnarray}
\left(\partial_{0}^{2}-\partial_{1}^{2}+m_{n}^{2}\right)
X^{\mu}_{n}(\tau, \sigma^{1})&=&0
\label{int12}
\end{eqnarray}
and the BC(s)
\begin{eqnarray}
\left(\partial_{1}X^{\mu}_{n}(\tau, \sigma^{1})-{B^{\mu}}_{\nu}
\partial_{0}X^{\nu}_{n}(\tau, \sigma^{1})\right)
|_{\sigma^{1}=0, \pi}&=&0~.
\label{bc}
\end{eqnarray}
The canonically conjugate momenta $\Pi_{n\mu}(\tau, \sigma^{1})$ to
$X^{\mu}_{n}(\tau, \sigma^{1})$ reads:
\begin{eqnarray}
\Pi_{n\mu}(\tau, \sigma^{1})&=&\eta_{\mu\nu}
\partial_{0}X^{\nu}_{-n}(\tau, \sigma^{1})-B_{\mu\nu}
\partial_{1}X^{\nu}_{-n}(\tau, \sigma^{1}).
\label{mom11}
\end{eqnarray}
Using $\Pi_{n\mu}(\tau, \sigma^{1})$, the above BC can be expressed
in terms of phase-space variables as:
\begin{eqnarray}
\left[{M^{\mu}}_{\rho}\partial_{1}X^{\rho}_{n}(\tau, \sigma^{1})
-{B^{\mu}}_{\rho}\Pi^{\rho}_{-n}(\tau, \sigma^{1})\right]
|_{\sigma^{1}=0, \pi}&=&0
\label{bcmom}
\end{eqnarray}
where, ${M^{\mu}}_{\rho}=({\delta^{\mu}}_{\rho}
-{B^{\mu}}_{\nu}{B^{\nu}}_{\rho})$~.

\noindent The solution to the equations of motion (\ref{int12})
compatible with the above BC(s) read \cite{chu}\footnote{We consider
the case in which the $B$--field takes the form 
$B_{\mu\nu}=\pmatrix{0&B\cr-B&0\cr}$.}
\begin{eqnarray}
X^{\mu}_{n}(\tau, \sigma^{1})&=& X^{\mu(0)}_{n}(\tau, \sigma^{1})
+X^{\mu(1)}_{n}(\tau, \sigma^{1})
\label{sol1}
\end{eqnarray}
where,
\begin{eqnarray}
X^{\mu(0)}_{n}(\tau, \sigma^{1})=&&
\left[x^{\mu}_{0n}\cos(\tilde{m}_{n}\tau)+p^{\mu}_{0n}
\frac{\sin(\tilde{m}_{n}\tau)}{\tilde{m}_{n}}\right]
\cosh(\tilde{m}_{n}B\sigma^{1})\nonumber\\
&&-{B^{\mu}}_{\nu}
\left[-p^{\nu}_{0n}\cos(\tilde{m}_{n}\tau)+\tilde{m}_{n}
x^{\nu}_{0n}\sin(\tilde{m}_{n}\tau)\right]
\frac{\sinh(\tilde{m}_{n}B\sigma^{1})}
{\tilde{m}_{n}B}
\label{sol1a}
\end{eqnarray}
is the ``zero mode" part (i.e. the modes with the lowest frequency) and
\begin{eqnarray}
X^{\mu(1)}_{n}(\tau, \sigma^{1})&=&\sum_{l\neq0}\frac{1}{\omega_{ln}}
e^{-i\omega_{ln}\tau}\left(ia^{\mu}_{ln}\cos(l\sigma^{1})+
\frac{\omega_{ln}}{l}{B^{\mu}}_{\nu}
a^{\nu}_{ln}\sin(l\sigma^{1})\right).
\label{sol1b}
\end{eqnarray}  
The constant $B$ is the eigen-value of the matrix ${B_{\mu}}^{\nu}$
and the frequencies are defined by: 
\begin{eqnarray}
\omega_{ln}&=&sgn(l)\sqrt{l^2+m_{n}^2}
\quad;\quad\tilde{m}_{n}=\frac{m_{n}}{\sqrt{1+B^2}}~.
\label{intsol1}
\end{eqnarray}  
Using $X^{\mu~\dagger}_{n}(\tau, \sigma^{1})
=X^{\mu}_{-n}(\tau, \sigma^{1})$, we find that the Fourier modes
of the massive strings satisfy the relations
\begin{eqnarray}
x^{\mu~\dagger}_{0n}&=&x^{\mu}_{0,-n}\quad;
\quad p^{\mu~\dagger}_{0n}=p^{\mu}_{0,-n}\quad;
\quad a^{\mu~\dagger}_{l, n}=a^{\mu}_{-l, -n}~.
\label{intsol1}
\end{eqnarray}  
The canonically conjugate momenta $\Pi^{(1)}_{n\mu}(\tau, \sigma^{1})$ 
to $X^{\mu(1)}_{n}(\tau, \sigma^{1})$ expressed in terms of the Fourier
modes read:
\begin{eqnarray}
\Pi^{\mu(1)}_{n}(\tau, \sigma^{1})&=&
\sum_{l>0}{M^{\mu}}_{\rho}\left[a^{\rho}_{l, -n}
e^{-i\omega_{ln}\tau}+a^{\rho}_{-l, -n}
e^{i\omega_{ln}\tau}\right]\cos(l\sigma^{1})\nonumber\\
&&+~i{B^{\mu}}_{\rho}\sum_{l>0}\left(\frac{l}{\omega_{ln}}
-\frac{\omega_{ln}}{l}\right)\left[a^{\rho}_{l, -n}
e^{-i\omega_{ln}\tau}-a^{\rho}_{-l, -n}
e^{i\omega_{ln}\tau}\right]\sin(l\sigma^{1}).
\label{canmom}
\end{eqnarray}  
Now we apply the FJ procedure to obtain the algebra
between the non-zero Fourier modes. To do this, we once
again write down the ``non-zero mode" part of the Lagrangian 
in a first order form
\begin{eqnarray}
L^{(1)}_{n}&=&\int_{0}^{\pi}
d\sigma^{1}~\Pi^{(1)}_{n\mu}\partial_{0}X^{\mu(1)}_{n}-H^{(1)}_{n}
\label{aa}
\end{eqnarray}  
which on substitution of the mode expansions (\ref{sol1b}) and (\ref{canmom})
read
\begin{eqnarray}
L^{(1)}_{n}=&&\frac{i\pi}{2}\eta_{\mu\kappa}\sum_{l>0}
\left[\frac{1}{\omega_{ln}}{Q^{\mu}}_{\rho(ln)}\left(a^{\rho}_{l, -n}
e^{-i\omega_{ln}\tau}+a^{\rho}_{-l, -n}
e^{i\omega_{ln}\tau}\right)\right.\nonumber\\
&&\left.~~~~~~~~~~~~~~~~~~~~~\times \partial_{0}\left(a^{\kappa}_{l, n}
e^{-i\omega_{ln}\tau}-a^{\kappa}_{-l, n}
e^{i\omega_{ln}\tau}\right)+...\right]-H^{(1)}_{n}
\label{aa1}
\end{eqnarray} 
where, ${Q^{\mu}}_{\rho(ln)}=({\delta^{\mu}}_{\rho}-\frac{\omega_{ln}^2}{l^2}
{B^{\mu}}_{\nu}{B^{\nu}}_{\rho})$ and the `...' represent terms which
do not play a role in the determination of the symplectic structure.
The explicit form of $H^{(1)}_{n}$ is also not required for obtaining
the PB(s) between the modes. 

\noindent As before, one can again read a set of variables $\xi^{l}_{n}$
and the corresponding canonical one-form $a_{nl}(\xi)$. The matrix
(\ref{mat1}) once again reads the same as (\ref{mat2}) with
the diagonal matrix $B$ being:
\begin{eqnarray}
B^{\mu\nu}_{lnl^{'}n^{'}}&=&-\frac{i\pi}{\omega_{ln}}sgn(l)
Q^{\mu\nu}_{(ln)}\delta_{n+n^{'},0}
~\delta_{l,l^{'}}~.
\label{matrix1}
\end{eqnarray} 
Now from the inverse of $f$ which reads the same as (\ref{mat6}), 
it is easy to read the PB(s) among the ``non-zero" Fourier modes 
using the FJ technique. They are:
\begin{eqnarray}
\{a^{\mu}_{ln}(\tau),~a^{\nu}_{l^{'},-n^{'}}(\tau)\}
&=&-\frac{i}{\pi}sgn(l)\omega_{ln}(Q^{-1}_{(ln)})^{\mu\nu}
~\delta_{n+n^{'},0}
~\delta_{l,l^{'}}~.
\label{abb}
\end{eqnarray}
The PB(s) among the zero modes $x^{\mu}_{0n}$ and $p^{\mu}_{0n}$
can be determined from the evolution equations (\ref{evo}, \ref{evo1})
and read:
\begin{eqnarray}
\{x^{\mu}_{0n},~p^{\nu}_{0n^{'}}\}&=&
\frac{\pi\tilde{m}_{n}\bar{B}}{\pi\tanh(\pi\tilde{m}_{n}\bar{B})}
(M^{-1})^{\mu\nu}~\delta_{n+n^{'},0}~.
\label{abc}
\end{eqnarray}
where $M_{\mu\nu}$ has been defined earlier.
Now since the BC(s) are valid on the boundaries, it is 
natural to demand that
\begin{eqnarray}
\{X^{\mu}_{n}(\tau, \sigma^{1}),~X^{\nu}_{n^{'}}(\tau, \sigma^{1'})\}
&=&0\quad; \quad for\quad \sigma^{1},~\sigma^{1'}~\in~(0, \pi)
\label{vanish}
\end{eqnarray} 
from which we get: 
\begin{eqnarray}
\{x^{\mu}_{0n},~x^{\nu}_{0n^{'}}\}&=&-(BM^{-1})^{\mu\nu}
~\delta_{n+n^{'},0}
\label{acc}
\end{eqnarray}
\begin{eqnarray}
\{p^{\mu}_{0n},~p^{\nu}_{0n^{'}}\}&=&-\tilde{m}_{n}^{2}
(BM^{-1})^{\mu\nu}~\delta_{n+n^{'},0}~.
\label{cc}
\end{eqnarray}
With the above results (\ref{abb}, \ref{abc}, \ref{acc}, \ref{cc}) in hand,
and after some lengthy calculation, we obtain the following PB(s):
\begin{eqnarray}
\{X^{\mu}_{n}(\tau, \sigma^{1}),~\Pi_{n^{'}\nu}(\tau, \sigma^{1'})\}
&=&{\delta^{\mu}}_{\nu}~\delta_{n+n^{'}, 0}
~\Delta_{+}(\sigma^{1}, \sigma^{1'})
\label{usualpb}
\end{eqnarray} 
\begin{eqnarray}
\{X^{\mu}_{n}(\tau, \sigma^{1}),~X^{\mu}_{n^{'}}(\tau, \sigma^{1'})\}
&=&\delta_{n+n^{'}, 0}~(BM^{-1})^{\mu\nu} \times
\left\{ 
\begin{array}{cl} 
1 ~, & \sigma^{1} = \sigma^{1'} =0 \cr
-1 ~, & \sigma^{1} =\sigma^{1'} =\pi \cr
0~, & \mbox{otherwise}. 
\end{array}
\right.
\label{cac}
\end{eqnarray} 
\begin{eqnarray} 
\label{PP}
\{\Pi_{n\mu}(\tau,\sigma^{1}), \Pi_{n^{'}\nu}(\tau,\sigma^{1'})\}&=&
\delta_{n+n^{'}, 0}~m_{n}^2 B_{\mu\nu} \times\left\{ 
\begin{array}{cl} 
+1 \;, & \sigma^{1} =\sigma^{1'}=0 \cr
-1 \;, & \sigma^{1} = \sigma^{1'}= \pi \cr
0~, \;& \mbox{otherwise}. 
\end{array}
\right.
\end{eqnarray}
which agrees with \cite{chu}. Note that in the $B\rightarrow 0$ limit,
the noncommutativity vanishes and the results are in conformity with
\cite{kamani}.

\section{Conclusions}
In this paper, we employ the Faddeev-Jackiw symplectic formalism
to study the problem of open free membrane and strings in pp-wave
background and background gauge field obtained by compactification
of interacting membrane. The starting point is the
solutions to the classical membrane(string) equations
of motion. It is then observed that one can find the PB(s)
among the Fourier modes first, using which the PB(s) among
the original variables can be obtained. This idea was first
proposed in \cite{ho} where the authors used the
time-independent symplectic form 
(\cite{cschu}, \cite{chu}) to fix the PB(s)
among the Fourier components. In this paper,
we follow a slightly different method (due to \cite{jian}) 
to obtain the symplectic structure
among the Fourier modes. The solutions of the membrane(string) equations
of motion are substituted into the Lagrangian and then integration
over the spatial variables is carried out to cast the Lagrangian
in a first order form involving the Fourier modes from which the PB(s)
(among the Fourier modes) can be easily read off. 
Finally, using this algebra we obtain a 
noncommutative phase-space structure. Our results agree
with the previous work in the literature \cite{chu}.
\section*{Acknowledgements }
The author would like
to thank the referee for useful comments.


\end{document}